\documentclass[aps,noshowpacs,preprint]{revtex4}
\usepackage{amssymb}
\usepackage{amsmath}
\usepackage{amsfonts}

\newtheorem{theorem}{Theorem}

\newtheorem{definition}[theorem]{Definition}

\newenvironment{proof}[1][Proof]{\noindent\textbf{#1.} }{\ \rule{0.5em}{0.5em}}
\begin{document}

\title{Multiplicative duality, $q$-triplet and $\left( \mu ,\nu ,q\right) $%
-relation\\
derived from the one-to-one correspondence \\
between the $\left( \mu ,\nu \right) $-multinomial coefficient and Tsallis
entropy $S_{q}$}
\author{Hiroki Suyari}
\email{suyari@faculty.chiba-u.jp, suyari@ieee.org}
\affiliation{Department of Information and Image Sciences, Chiba University, Chiba
263-8522, Japan}
\author{Tatsuaki Wada}
\email{wada@mx.ibaraki.ac.jp}
\affiliation{Department of Electrical and Electronic Engineering, Ibaraki University,
Hitachi, Ibaraki 316-8511, Japan}
\pacs{PACS number}

\begin{abstract}
We derive the multiplicative duality \textquotedblleft $q\leftrightarrow 1/q$%
\textquotedblright\ and other typical mathematical structures as the special
cases of the $\left( \mu ,\nu ,q\right) $-relation behind Tsallis statistics
by means of the $\left( \mu ,\nu \right) $-multinomial coefficient. Recently
the additive duality \textquotedblleft $q\leftrightarrow 2-q$%
\textquotedblright\ in Tsallis statistics is derived in the form of the
one-to-one correspondence between the $q$-multinomial coefficient and
Tsallis entropy. A slight generalization of this correspondence for the
multiplicative duality requires the $\left( \mu ,\nu \right) $-multinomial
coefficient as a generalization of the $q$-multinomial coefficient. This
combinatorial formalism provides us with the one-to-one correspondence
between the $\left( \mu ,\nu \right) $-multinomial coefficient and Tsallis
entropy $S_{q}$, which determines a concrete relation among three parameters 
$\mu ,\nu $ and $q$, i.e., $\nu \left( 1-\mu \right) +1=q$ which is called
\textquotedblleft $\left( \mu ,\nu ,q\right) $-relation\textquotedblright\
in this paper. As special cases of the $\left( \mu ,\nu ,q\right) $%
-relation, the additive duality and the multiplicative duality are recovered
when $\nu =1$ and $\nu =q$, respectively. As other special cases, when $\nu
=2-q$, a set of three parameters $\left( \mu ,\nu ,q\right) $ is identified
with the $q$-triplet $\left( q_{\text{sen}},q_{\text{rel}},q_{\text{stat}%
}\right) $ recently conjectured by Tsallis. Moreover, when $\nu =1/q$, the
relation $1/\left( 1-q_{\text{sen}}\right) =1/\alpha _{\min }-1/\alpha
_{\max }$ in the multifractal singularity spectrum $f\left( \alpha \right) $
is recovered by means of the $\left( \mu ,\nu ,q\right) $-relation.
\end{abstract}

\date{\today }
\maketitle

\section{Introduction}

In the last two decades the so-called Tsallis statistics or $q$-statistics
has been introduced \cite{Ts88} and studied as a generalization of
Boltzmann-Gibbs statistics with many applications to complex systems \cite%
{AO01}\cite{GT04}, whose information measure is given by%
\begin{equation}
S_{q}\left( {p_{1},\ldots ,p_{n}}\right) ={\frac{{{1-\sum\limits_{i=1}^{n}{%
p_{i}^{q}}}}}{{q-1}}}  \label{Tsallis entropy}
\end{equation}%
where $p_{i}$ is a probability of $i$th state and $q$ is a real parameter.
This generalized entropy $S_{q}$ is nowadays called \textit{Tsallis entropy}
which recovers Boltzmann-Gibbs-Shannon entropy $S_{1}$ when $q\rightarrow 1$%
. The above entropic form (\ref{Tsallis entropy}) was first given in \cite%
{HC67} and \cite{Da70} from a mathematical motivation, but in 1988 \cite%
{Ts88} Tsallis first applied the above form (\ref{Tsallis entropy}) to a
generalization of Boltzmann-Gibbs statistics for nonequilibrium systems
through the maximum entropy principle (MaxEnt for short) along the lines of
Jaynes approach \cite{Ja57}. Since then, many applications of (\ref{Tsallis
entropy}) to the studies of complex systems with power-law behaviors have
been presented using the MaxEnt as a main approach \cite{TMP98}. In fact,
the $q$-exponential function appeared in the MaxEnt plays a crucial role in
the formalism and applications \cite{AO01}\cite{GT04}.

For all many applications of the MaxEnt for Tsallis entropy (\ref{Tsallis
entropy}), there have been missing a combinatorial consideration in Tsallis
statistics until recently \cite{Su04b}, whose ideas originate from
Boltzmann's pioneering work \cite{Bo1877} (See \cite{Ni05} for the
comprehensive review). By means of the $q$-product uniquely determined by
the $q$-exponential function \cite{NMW03}\cite{Bo03} as the $q$-exponential
law, the one-to-one correspondence between the $q$-multinomial coefficient
and Tsallis entropy is obtained as follows \cite{Su04b}: for $%
n=\sum_{i=1}^{k}n_{i}$ and $n_{i}\in \mathbb{N}$ if $q\neq 2$, 
\begin{equation}
\ln _{q}\left[ 
\begin{array}{ccc}
& n &  \\ 
n_{1} & \cdots & n_{k}%
\end{array}%
\right] _{q}\simeq \dfrac{n^{2-q}}{2-q}\cdot S_{2-q}\left( \dfrac{n_{1}}{n}%
,\cdots ,\dfrac{n_{k}}{n}\right)  \label{q-correspondence}
\end{equation}%
where $\left[ 
\begin{array}{ccc}
& n &  \\ 
n_{1} & \cdots & n_{k}%
\end{array}%
\right] _{q}$ is the $q$-multinomial coefficient and $\ln _{q}$ is the $q$%
-logarithm. The above correspondence (\ref{q-correspondence}) obviously
recovers the well-known correspondence:%
\begin{equation}
\ln \left[ 
\begin{array}{ccc}
& n &  \\ 
n_{1} & \cdots & n_{k}%
\end{array}%
\right] \simeq n\cdot S_{1}\left( \dfrac{n_{1}}{n},\cdots ,\dfrac{n_{k}}{n}%
\right)
\end{equation}%
when $q\rightarrow 1$. Moreover, the \textit{additive duality}
\textquotedblleft $q\leftrightarrow 2-q$\textquotedblright\ in Tsallis
statistics is revealed in (\ref{q-correspondence}). In the MaxEnt formalism
for Tsallis entropy, two kinds of dualities \textquotedblleft $%
q\leftrightarrow 2-q$\textquotedblright\ and \textquotedblleft $%
q\leftrightarrow 1/q$\textquotedblright\ have been observed and discussed 
\cite{TMP98}\cite{Ra99}\cite{Na02}\cite{Na04}\cite{WS05}, but in the
combinatorial formalism the \textit{multiplicative duality}
\textquotedblleft $q\leftrightarrow 1/q$\textquotedblright\ is still missing.

In this paper, we derive the multiplicative duality \textquotedblleft $%
q\leftrightarrow 1/q$\textquotedblright\ along the lines of the above
correspondence (\ref{q-correspondence}), which introduces the $\left( \mu
,\nu \right) $-factorial as a generalization of the $q$-factorial. We apply
the $\left( \mu ,\nu \right) $-factorial to the formulation of the $\left(
\mu ,\nu \right) $-multinomial coefficient and the $\left( \mu ,\nu \right) $%
-Stirling's formula, which results in the following correspondence: for $%
n=\sum_{i=1}^{k}n_{i}$ and $n_{i}\in \mathbb{N}$ if $q,\nu \neq 0$, 
\begin{equation}
\frac{1}{\nu }\ln _{\mu }\left[ 
\begin{array}{ccc}
& n &  \\ 
n_{1} & \cdots  & n_{k}%
\end{array}%
\right] _{\left( \mu ,\nu \right) }\simeq \dfrac{n^{q}}{q}\cdot S_{q}\left( 
\dfrac{n_{1}}{n},\cdots ,\dfrac{n_{k}}{n}\right)   \label{mnu-correspondence}
\end{equation}%
where $\left[ 
\begin{array}{ccc}
& n &  \\ 
n_{1} & \cdots  & n_{k}%
\end{array}%
\right] _{\left( \mu ,\nu \right) }$ is the $\left( \mu ,\nu \right) $%
-multinomial coefficient and three parameters $\mu ,\nu ,q$ satisfy the
relation: 
\begin{equation}
\nu \left( 1-\mu \right) +1=q  \label{parameter-relation}
\end{equation}%
which is called \textquotedblleft $\left( \mu ,\nu ,q\right) $\textit{%
-relation}\textquotedblright\ throughout the paper.

Using the additive duality \textquotedblleft $q\leftrightarrow 2-q$%
\textquotedblright\ in (\ref{q-correspondence}), (\ref{q-correspondence}) is
rewritten by%
\begin{equation}
\ln _{2-q}\left[ 
\begin{array}{ccc}
& n &  \\ 
n_{1} & \cdots & n_{k}%
\end{array}%
\right] _{2-q}\simeq \dfrac{n^{q}}{q}\cdot S_{q}\left( \dfrac{n_{1}}{n}%
,\cdots ,\dfrac{n_{k}}{n}\right) .  \label{rev_q-correspondence}
\end{equation}%
Hence the above generalized correspondence (\ref{mnu-correspondence}) is
found to recover (\ref{rev_q-correspondence}) when $\mu =2-q$ and $\nu =1$.
As will be shown later,%
\begin{equation}
\left[ 
\begin{array}{ccc}
& n &  \\ 
n_{1} & \cdots & n_{k}%
\end{array}%
\right] _{\left( \mu ,1\right) }=\left[ 
\begin{array}{ccc}
& n &  \\ 
n_{1} & \cdots & n_{k}%
\end{array}%
\right] _{\mu }.
\end{equation}

The $\left( \mu ,\nu ,q\right) $-relation (\ref{parameter-relation}) among
three parameters $\mu ,\nu ,q$ yields the additive duality \textquotedblleft 
$q\leftrightarrow 2-q$\textquotedblright\ when $\nu =1$ and the
multiplicative duality \textquotedblleft $q\leftrightarrow 1/q$%
\textquotedblright\ when $\nu =q$, respectively. As other special cases of
the $\left( \mu ,\nu ,q\right) $-relation, when $\nu =2-q$, it is shown that
the $q$\textit{-triplet} $\left( q_{\text{sen}},q_{\text{rel}},q_{\text{stat}%
}\right) $ recently conjectured by Tsallis \cite{Ts04PA} is identified with\
the $\left( \mu ,\nu ,q\right) $-relation (\ref{parameter-relation}) in the
following sense: 
\begin{equation}
\mu =\frac{1}{q_{\text{sen}}},\text{\quad }\nu =\frac{1}{q_{\text{rel}}},%
\text{\quad }q=q_{\text{stat}}.  \label{relation to q-troplet}
\end{equation}%
Moreover, when $\nu =1/q$, the relation \cite{LT98}:%
\begin{equation}
\frac{1}{1-q_{\text{sen}}}=\frac{1}{\alpha _{\min }}-\frac{1}{\alpha _{\max }%
}
\end{equation}%
in the multifractal singularity spectrum $f\left( \alpha \right) $ is
recovered by means of the $\left( \mu ,\nu ,q\right) $-relation in the
following sense: 
\begin{equation}
\mu =q_{\text{sen}},\text{\quad }\nu =\frac{1}{\alpha _{\max }},\text{\quad }%
q=\alpha _{\max }  \label{relation to multifractal}
\end{equation}%
with $\alpha _{\max }-\alpha _{\min }=1$. The above new results are derived
in detail in the following sections.

This paper consists of the 5 sections including this introduction. In the
next section, we briefly review the fundamental formulas such as the $q$%
-product, the $q$-factorial, the $q$-multinomial coefficient and the $q$%
-Stirling's formula which are applied to the derivation of (\ref%
{q-correspondence}). In Section III, the correspondence (\ref%
{q-correspondence}) is modified to derive the multiplicative duality
\textquotedblleft $q\leftrightarrow 1/q$\textquotedblright\ in our
combinatorial formalism. In this derivation, a slight generalization of the $%
q$-factorial is required, which is called \textquotedblleft $\left( \mu ,\nu
\right) $-factorial\textquotedblright . As similarly as Section II, we
formulate the $\left( \mu ,\nu \right) $-multinomial coefficient and the $%
\left( \mu ,\nu \right) $-Stirling's formula based on the $\left( \mu ,\nu
\right) $-factorial, and apply them to finding the generalized
correspondence (\ref{mnu-correspondence}). In Section IV, we derive the
additive duality and the multiplicative duality as special cases of (\ref%
{mnu-correspondence}). Moreover, when $\nu =2-q$ and $\nu =1/q$, each
interpretation of the $\left( \mu ,\nu ,q\right) $-relation shown in (\ref%
{relation to q-troplet}) and (\ref{relation to multifractal}) is
respectively presented. The final section is devoted to our conclusion.

\section{Additive duality derived from the $q$-multinomial coefficient}

The MaxEnt for Boltzmann-Gibbs-Shannon entropy $S_{1}$ yields the
exponential function $\exp \left( x\right) $ which is well known to be
characterized by the linear differential function $dy/dx=y$. In parallel
with this, the MaxEnt for Tsallis entropy $S_{q}$ yields a generalization of
the exponential function $\exp _{q}\left( x\right) $ \cite{TMP98}\cite{WS05b}%
\cite{Su06} which is characterized by the nonlinear differential function $%
dy/dx=y^{q}$ \cite{Ts04PD}\cite{Su06b}. In Tsallis statistics, the
fundamental functions are the $q$\textit{-logarithm} $\ln _{q}x$ and the $q$%
\textit{-exponential} $\exp _{q}\left( x\right) $, respectively defined as
follows:

\begin{definition}
($q$\textit{-logarithm, }$q$\textit{-exponential}) The $q$\textit{-logarithm}
$\ln _{q}x:\mathbb{R}^{+}\rightarrow \mathbb{R}$ and the $q$\textit{%
-exponential} $\exp _{q}\left( x\right) :\mathbb{R}\rightarrow \mathbb{R}$
are defined by%
\begin{equation}
\ln _{q}x:=\frac{x^{1-q}-1}{1-q},  \label{q-logarithm}
\end{equation}%
\begin{equation}
\exp _{q}\left( x\right) :=\left\{ 
\begin{array}{ll}
\left[ 1+\left( 1-q\right) x\right] ^{\frac{1}{1-q}} & \text{if }1+\left(
1-q\right) x>0, \\ 
0 & \text{otherwise.}%
\end{array}%
\right.  \label{q-exponential}
\end{equation}
\end{definition}

Then a new product $\otimes _{q}$ to satisfy the following identities as the 
$q$\textit{-exponential law} is introduced.%
\begin{align}
\ln _{q}\left( x\otimes _{q}y\right) & =\ln _{q}x+\ln _{q}y,
\label{requirement1} \\
\exp _{q}\left( x\right) \otimes _{q}\exp _{q}\left( y\right) & =\exp
_{q}\left( x+y\right) .  \label{requirement2}
\end{align}%
For this purpose, the new multiplication operation $\otimes _{q}$ is
introduced in \cite{NMW03}\cite{Bo03}. The concrete forms of the $q$%
-logarithm and $q$-exponential are given in (\ref{q-logarithm}) and (\ref%
{q-exponential}), so that the above requirement (\ref{requirement1}) or (\ref%
{requirement2}) as the $q$-exponential law leads to the definition of $%
\otimes _{q}$ between two positive numbers.

\begin{definition}
($q$\textit{-product}) For $x,y\in \mathbb{R}^{+}$, the $q$-\textit{product} 
$\otimes _{q}$ is defined by%
\begin{equation}
x\otimes _{q}y:=\left\{ 
\begin{array}{ll}
\left[ x^{1-q}+y^{1-q}-1\right] ^{\frac{1}{1-q}}, & \text{if }%
x>0,\,y>0,\,x^{1-q}+y^{1-q}-1>0, \\ 
0, & \text{otherwise.}%
\end{array}%
\right.  \label{def of q-product}
\end{equation}
\end{definition}

The $q$-\textit{product} recovers the usual product such that $\underset{%
q\rightarrow 1}{\lim }\left( x\otimes _{q}y\right) =xy$. The fundamental
properties of the $q$-product $\otimes _{q}$ are almost the same as the
usual product, but 
\begin{equation}
a\left( x\otimes _{q}y\right) \neq \left( ax\right) \otimes _{q}y\quad
\left( a,x,y\in \mathbb{R}\right) .
\end{equation}%
The other properties of the $q$-\textit{product} are available in \cite%
{NMW03}\cite{Bo03}.

By means of the $q$-product (\ref{def of q-product}), the $q$-factorial is
naturally defined in the following form \cite{Su04b}.

\begin{definition}
($q$-factorial) For a natural number $n\in \mathbb{N}$ and $q\in \mathbb{R}%
^{+}$, the $q$-factorial $n!_{q}$ is defined by%
\begin{equation}
n!_{q}:=1\otimes _{q}\cdots \otimes _{q}n.  \label{def of q-kaijyo}
\end{equation}
\end{definition}

Thus, we concretely compute the $q$-Stirling's formula.

\begin{theorem}
($q$-Stirling's formula) Let $n!_{q}$ be the $q$-factorial defined by (\ref%
{def of q-kaijyo}). The rough $q$-Stirling's formula $\ln _{q}\left(
n!_{q}\right) $ is computed as follows:%
\begin{equation}
\ln _{q}\left( n!_{q}\right) =\left\{ 
\begin{array}{ll}
\dfrac{n\ln _{q}n-n}{2-q}+O\left( \ln _{q}n\right) \quad & \text{if}\quad
q\neq 2, \\ 
n-\ln n+O\left( 1\right) & \text{if}\quad q=2.%
\end{array}%
\right.  \label{rough q-Stirling}
\end{equation}
\end{theorem}

The above rough $q$-Stirling's formula is obtained by the approximation:%
\begin{equation}
\ln _{q}\left( n!_{q}\right) =\sum_{k=1}^{n}\ln _{q}k\simeq \int_{1}^{n}\ln
_{q}xdx.
\end{equation}%
The rigorous derivation of the $q$-Stirling's formula is given in \cite%
{Su04b}.

Similarly as for the $q$-product, $q$\textit{-ratio} is introduced from the
requirements: 
\begin{align}
\ln _{q}\left( x\oslash _{q}y\right) & =\ln _{q}x-\ln _{q}y, \\
\exp _{q}\left( x\right) \oslash _{q}\exp _{q}\left( y\right) & =\exp
_{q}\left( x-y\right) .
\end{align}%
Then we define the $q$-ratio as follows.

\begin{definition}
($q$\textit{-ratio}) For $x,y\in \mathbb{R}^{+}$, the inverse operation to
the $q$-product is defined by 
\begin{equation}
x\oslash _{q}y:=\left\{ 
\begin{array}{ll}
\left[ x^{1-q}-y^{1-q}+1\right] ^{\frac{1}{1-q}}, & \text{if }%
x>0,\,y>0,\,x^{1-q}-y^{1-q}+1>0, \\ 
0, & \text{otherwise}%
\end{array}%
\right.
\end{equation}%
which is called $q$\textit{-ratio} in \cite{Bo03}.
\end{definition}

The $q$-product, $q$-factorial and $q$-ratio are applied to the definition
of the $q$-multinomial coefficient \cite{Su04b}.

\begin{definition}
($q$-multinomial coefficient) For $n=\sum_{i=1}^{k}n_{i}$ and $n_{i}\in 
\mathbb{N\,}\left( i=1,\cdots ,k\right) ,$ the $q$-multinomial coefficient
is defined by%
\begin{equation}
\left[ 
\begin{array}{ccc}
& n &  \\ 
n_{1} & \cdots & n_{k}%
\end{array}%
\right] _{q}:=\left( n!_{q}\right) \oslash _{q}\left[ \left(
n_{1}!_{q}\right) \otimes _{q}\cdots \otimes _{q}\left( n_{k}!_{q}\right) %
\right] .  \label{def of q-multinomial coefficient}
\end{equation}
\end{definition}

From the definition (\ref{def of q-multinomial coefficient}), it is clear
that%
\begin{equation}
\underset{q\rightarrow 1}{\lim }\left[ 
\begin{array}{ccc}
& n &  \\ 
n_{1} & \cdots & n_{k}%
\end{array}%
\right] _{q}=\left[ 
\begin{array}{ccc}
& n &  \\ 
n_{1} & \cdots & n_{k}%
\end{array}%
\right] =\frac{n!}{n_{1}!\cdots n_{k}!}.
\end{equation}%
Throughout the present paper, we consider the $q$-logarithm of the $q$%
-multinomial coefficient to be given by 
\begin{equation}
\ln _{q}\left[ 
\begin{array}{ccc}
& n &  \\ 
n_{1} & \cdots & n_{k}%
\end{array}%
\right] _{q}=\ln _{q}\left( n!_{q}\right) -\ln _{q}\left( n_{1}!_{q}\right)
\cdots -\ln _{q}\left( n_{k}!_{q}\right) .  \label{lnq-q-multinomial}
\end{equation}

Based on these fundamental formulas, we obtain the one-to-one correspondence
(\ref{q-correspondence}) between the $q$-multinomial coefficient and Tsallis
entropy as follows \cite{Su04b}.

\begin{theorem}
When $n\in \mathbb{N}$ is sufficiently large, the $q$-logarithm of the $q$%
-multinomial coefficient coincides with Tsallis entropy (\ref{Tsallis
entropy}) in the following correspondence:%
\begin{equation}
\ln _{q}\left[ 
\begin{array}{ccc}
& n &  \\ 
n_{1} & \cdots & n_{k}%
\end{array}%
\right] _{q}\simeq \left\{ 
\begin{array}{ll}
\dfrac{n^{2-q}}{2-q}\cdot S_{2-q}\left( \dfrac{n_{1}}{n},\cdots ,\dfrac{n_{k}%
}{n}\right) & \text{if}\quad q>0,\,\,q\neq 2 \\ 
-S_{1}\left( n\right) +\sum\limits_{i=1}^{k}S_{1}\left( n_{i}\right) & \text{%
if}\quad q=2%
\end{array}%
\right.  \label{2-q-correspondence}
\end{equation}%
where $S_{q}$ is Tsallis entropy (\ref{Tsallis entropy}) and $S_{1}\left(
n\right) :=\ln n.$
\end{theorem}

Straightforward computation of the left side of (\ref{2-q-correspondence})
by means of the $q$-Stirling's formula (\ref{rough q-Stirling}) yields the
above result (\ref{2-q-correspondence}). (See \cite{Su04b} for the proof.)

Clearly the additive duality \textquotedblleft $q\leftrightarrow 2-q$%
\textquotedblright\ is appeared in the above one-to-one correspondence (\ref%
{2-q-correspondence}). In the following sections these fundamental formulas
are generalized for the derivation of the multiplicative duality
\textquotedblleft $q\leftrightarrow 1/q$\textquotedblright\ in the similar
correspondence as (\ref{2-q-correspondence}).

\section{One-to-one correspondence between the $\left( \protect\mu ,\protect%
\nu \right) $-multinomial coefficient and Tsallis entropy}

In this section the correspondence (\ref{2-q-correspondence}) is generalized
for the multiplicative duality \textquotedblleft $q\leftrightarrow 1/q$%
\textquotedblright . For this purpose, replace $q$ in (\ref%
{2-q-correspondence}) by $1/q$ at first. Then we obtain 
\begin{equation}
\ln _{\frac{1}{q}}\left[ 
\begin{array}{ccc}
& n &  \\ 
n_{1} & \cdots & n_{k}%
\end{array}%
\right] _{\frac{1}{q}}\simeq \frac{n^{2-\frac{1}{q}}}{2-\frac{1}{q}}\cdot
S_{2-\frac{1}{q}}\left( \frac{n_{1}}{n},\cdots ,\frac{n_{k}}{n}\right)
\label{*2-q-correspondence}
\end{equation}%
where we consider the case $q>0$ and $q\neq 1/2$ only. The left side of (\ref%
{*2-q-correspondence}) is computed as 
\begin{equation}
\ln _{\frac{1}{q}}\left[ 
\begin{array}{ccc}
& n &  \\ 
n_{1} & \cdots & n_{k}%
\end{array}%
\right] _{\frac{1}{q}}=\ln _{\frac{1}{q}}\left( n!_{\frac{1}{q}}\right) -\ln
_{\frac{1}{q}}\left( n_{1}!_{\frac{1}{q}}\right) \cdots -\ln _{\frac{1}{q}%
}\left( n_{k}!_{\frac{1}{q}}\right) .  \label{1/q-logarithm}
\end{equation}

\textit{Using this formula (\ref{1/q-logarithm}), we will represent the left
side of (\ref{*2-q-correspondence}) by means of the forms such as }$\ln _{q}$%
\textit{\ or }$\ln _{2-q}$\textit{\ to find the multiplicative duality.} The
important relation for this purpose is the following identity:%
\begin{equation}
\ln _{\frac{1}{q}}\left( \frac{1}{x^{q}}\right) =-q\ln _{q}x.
\label{q-log-koushiki}
\end{equation}%
Each term $\ln _{\frac{1}{q}}\left( n!_{\frac{1}{q}}\right) $ on the right
side of (\ref{1/q-logarithm}) is equal to%
\begin{equation}
\ln _{\frac{1}{q}}\left( n!_{\frac{1}{q}}\right) =\ln _{\frac{1}{q}}\left( 
\frac{1}{\left( n!_{\frac{1}{q}}\right) ^{-1}}\right) .  \label{ln1/q-1}
\end{equation}%
$\left( n!_{\frac{1}{q}}\right) ^{-1}$ is expanded in accordance with the
definition of the $q$-product (\ref{def of q-product}).%
\begin{align}
\left( n!_{\frac{1}{q}}\right) ^{-1}& =\left( 1\otimes _{\frac{1}{q}}\cdots
\otimes _{\frac{1}{q}}n\right) ^{-1}=\left[ 1^{1-\frac{1}{q}}+2^{1-\frac{1}{q%
}}+\cdots +n^{1-\frac{1}{q}}-\left( n-1\right) \right] ^{\frac{-1}{1-\frac{1%
}{q}}} \\
& =\left[ \left( \left( \frac{1}{1}\right) ^{\frac{1}{q}}\right)
^{1-q}+\left( \left( \frac{1}{2}\right) ^{\frac{1}{q}}\right) ^{1-q}+\cdots
+\left( \left( \frac{1}{n}\right) ^{\frac{1}{q}}\right) ^{1-q}-\left(
n-1\right) \right] ^{\frac{q}{1-q}} \\
& =\left[ \left( \frac{1}{1}\right) ^{\frac{1}{q}}\otimes _{q}\left( \frac{1%
}{2}\right) ^{\frac{1}{q}}\otimes _{q}\cdots \otimes _{q}\left( \frac{1}{n}%
\right) ^{\frac{1}{q}}\right] ^{q}  \label{-1expandby q-product}
\end{align}%
Then $\ln _{\frac{1}{q}}\left( n!_{\frac{1}{q}}\right) $ is given by%
\begin{align}
\ln _{\frac{1}{q}}\left( n!_{\frac{1}{q}}\right) & =\ln _{\frac{1}{q}}\left( 
\frac{1}{\left[ \left( \frac{1}{1}\right) ^{\frac{1}{q}}\otimes _{q}\left( 
\frac{1}{2}\right) ^{\frac{1}{q}}\otimes _{q}\cdots \otimes _{q}\left( \frac{%
1}{n}\right) ^{\frac{1}{q}}\right] ^{q}}\right) \quad \left( \because \text{(%
\ref{ln1/q-1}) and (\ref{-1expandby q-product})}\right) \\
& =-q\ln _{q}\left[ \left( \frac{1}{1}\right) ^{\frac{1}{q}}\otimes
_{q}\left( \frac{1}{2}\right) ^{\frac{1}{q}}\otimes _{q}\cdots \otimes
_{q}\left( \frac{1}{n}\right) ^{\frac{1}{q}}\right] \quad \left( \because 
\text{(\ref{q-log-koushiki})}\right) \\
& =-q\sum_{j=1}^{n}\ln _{q}j^{-\frac{1}{q}}.  \label{lnq_expandby q-product}
\end{align}%
Thus, substitution of (\ref{lnq_expandby q-product}) to (\ref{1/q-logarithm}%
) yields%
\begin{equation}
\ln _{\frac{1}{q}}\left[ 
\begin{array}{ccc}
& n &  \\ 
n_{1} & \cdots & n_{k}%
\end{array}%
\right] _{\frac{1}{q}}=q\left( -\sum_{j=1}^{n}\ln _{q}j^{-\frac{1}{q}%
}+\sum_{j_{1}=1}^{n_{1}}\ln _{q}j_{1}^{-\frac{1}{q}}+\cdots
+\sum_{j_{k}=1}^{n_{k}}\ln _{q}j_{k}^{-\frac{1}{q}}\right) .  \label{henkei0}
\end{equation}%
Applying the general formula:%
\begin{equation}
-\ln _{q}j^{-\frac{1}{q}}=\ln _{2-q}j^{\frac{1}{q}}
\end{equation}%
to the above result (\ref{henkei0}), we have%
\begin{align}
\frac{1}{q}\ln _{\frac{1}{q}}\left[ 
\begin{array}{ccc}
& n &  \\ 
n_{1} & \cdots & n_{k}%
\end{array}%
\right] _{\frac{1}{q}}& =\sum_{j=1}^{n}\ln _{2-q}j^{\frac{1}{q}%
}-\sum_{j_{1}=1}^{n_{1}}\ln _{2-q}j_{1}^{\frac{1}{q}}-\cdots
-\sum_{j_{k}=1}^{n_{k}}\ln _{2-q}j_{k}^{\frac{1}{q}} \\
& =\ln _{2-q}\left[ \left( 1^{\frac{1}{q}}\otimes _{2-q}2^{\frac{1}{q}%
}\otimes _{2-q}\cdots \otimes _{2-q}n^{\frac{1}{q}}\right) \right.  \notag \\
& \oslash _{2-q}\left( 1^{\frac{1}{q}}\otimes _{2-q}2^{\frac{1}{q}}\otimes
_{2-q}\cdots \otimes _{2-q}n_{1}^{\frac{1}{q}}\right)  \notag \\
& \cdots  \notag \\
& \left. \oslash _{2-q}\left( 1^{\frac{1}{q}}\otimes _{2-q}2^{\frac{1}{q}%
}\otimes _{2-q}\cdots \otimes _{2-q}n_{k}^{\frac{1}{q}}\right) \right] .
\label{henkei1}
\end{align}

On the other hand, from (\ref{lnq-q-multinomial}) the $q$-logarithm of the $%
q $-multinomial coefficient is given by 
\begin{align}
\ln _{q}\left[ 
\begin{array}{ccc}
& n &  \\ 
n_{1} & \cdots & n_{k}%
\end{array}%
\right] _{q}& =\sum_{j=1}^{n}\ln _{q}j-\sum_{j_{1}=1}^{n_{1}}\ln
_{q}j_{1}-\cdots -\sum_{j_{k}=1}^{n_{k}}\ln _{q}j_{k}  \notag \\
& =\ln _{q}\left[ \left( 1\otimes _{q}2\otimes _{q}\cdots \otimes
_{q}n\right) \right.  \notag \\
& \oslash _{q}\left( 1\otimes _{q}2\otimes _{q}\cdots \otimes
_{q}n_{1}\right)  \notag \\
& \cdots  \notag \\
& \left. \oslash _{q}\left( 1\otimes _{q}2\otimes _{q}\cdots \otimes
_{q}n_{k}\right) \right] .  \label{henkei2}
\end{align}%
Comparing the argument of $\ln _{2-q}$ on the right side of (\ref{henkei1})
with that of $\ln _{q}$ on (\ref{henkei2}), a generalization of the $q$%
-factorial (\ref{def of q-kaijyo}) is found to be required for our purpose.

\begin{definition}
($\left( \mu ,\nu \right) $-factorial) For a natural number $n\in \mathbb{N}$
and $\mu ,\nu \in \mathbb{R}$, the $\left( \mu ,\nu \right) $-factorial $%
n!_{\left( \mu ,\nu \right) }$ is defined by%
\begin{equation}
n!_{\left( \mu ,\nu \right) }:=1^{\nu }\otimes _{\mu }2^{\nu }\otimes _{\mu
}\cdots \otimes _{\mu }n^{\nu }.  \label{def of munu-kaijyo}
\end{equation}%
where $\nu \neq 0$.
\end{definition}

Clearly when $\mu =q,\nu =1$ the $q$-factorial (\ref{def of q-kaijyo}) is
recovered. 
\begin{equation}
n!_{q}=n!_{\left( q,1\right) }
\end{equation}%
Moreover, when $\mu =1$, the $\left( \mu ,\nu \right) $-factorial $%
n!_{\left( \mu ,\nu \right) }$ is equal to $\left( n!\right) ^{\nu }$
because the $\mu $-product recovers the usual product. 
\begin{equation}
n!_{\left( 1,\nu \right) }=1^{\nu }2^{\nu }\cdots n^{\nu }=\left( n!\right)
^{\nu }
\end{equation}%
Thus, throughout the paper we consider the case $\mu \neq 1$ only.

Using the $\left( \mu ,\nu \right) $-factorial, we have 
\begin{equation}
1^{\frac{1}{q}}\otimes _{2-q}2^{\frac{1}{q}}\otimes _{2-q}\cdots \otimes
_{2-q}n^{\frac{1}{q}}=n!_{\left( 2-q,\frac{1}{q}\right) },
\end{equation}%
so that (\ref{henkei1}) is rewritten by means of the $\left( \mu ,\nu
\right) $-factorial.%
\begin{equation}
\frac{1}{q}\ln _{\frac{1}{q}}\left[ 
\begin{array}{ccc}
& n &  \\ 
n_{1} & \cdots  & n_{k}%
\end{array}%
\right] _{\frac{1}{q}}=\ln _{2-q}\left( n!_{\left( 2-q,\frac{1}{q}\right)
}\oslash _{2-q}n_{1}!_{\left( 2-q,\frac{1}{q}\right) }\cdots \oslash
_{2-q}n_{k}!_{\left( 2-q,\frac{1}{q}\right) }\right) 
\label{-1expandby munu-product}
\end{equation}%
Then we define the form of the argument of $\ln _{2-q}$ on the right side of
(\ref{-1expandby munu-product}) by the $\left( \mu ,\nu \right) $%
-multinomial coefficient as a generalization of the $q$-multinomial
coefficient (\ref{def of q-multinomial coefficient}).

\begin{definition}
($\left( \mu ,\nu \right) $-multinomial coefficient) For $%
n=\sum_{i=1}^{k}n_{i}$ and $n_{i}\in \mathbb{N\,}\left( i=1,\cdots ,k\right)
,$ the $\left( \mu ,\nu \right) $-multinomial coefficient is defined by%
\begin{equation}
\left[ 
\begin{array}{ccc}
& n &  \\ 
n_{1} & \cdots & n_{k}%
\end{array}%
\right] _{\left( \mu ,\nu \right) }:=\left( n!_{\left( \mu ,\nu \right)
}\right) \oslash _{\mu }\left[ \left( n_{1}!_{\left( \mu ,\nu \right)
}\right) \otimes _{\mu }\cdots \otimes _{\mu }\left( n_{k}!_{\left( \mu ,\nu
\right) }\right) \right] .
\end{equation}%
where $n!_{\left( \mu ,\nu \right) }$ is the $\left( \mu ,\nu \right) $%
-factorial defined in (\ref{def of munu-kaijyo}).
\end{definition}

Clearly when $\mu =q$ and $\nu =1$ the $q$-multinomial coefficient (\ref{def
of q-multinomial coefficient}) is recovered. 
\begin{equation}
\left[ 
\begin{array}{ccc}
& n &  \\ 
n_{1} & \cdots  & n_{k}%
\end{array}%
\right] _{q}=\left[ 
\begin{array}{ccc}
& n &  \\ 
n_{1} & \cdots  & n_{k}%
\end{array}%
\right] _{\left( q,1\right) }
\end{equation}%
Using the $\left( \mu ,\nu \right) $-multinomial coefficient, (\ref%
{-1expandby munu-product}) becomes%
\begin{equation}
\frac{1}{q}\ln _{\frac{1}{q}}\left[ 
\begin{array}{ccc}
& n &  \\ 
n_{1} & \cdots  & n_{k}%
\end{array}%
\right] _{\left( \frac{1}{q},1\right) }=\ln _{2-q}\left[ 
\begin{array}{ccc}
& n &  \\ 
n_{1} & \cdots  & n_{k}%
\end{array}%
\right] _{\left( 2-q,\frac{1}{q}\right) }
\end{equation}%
Moreover, the $\left( \mu ,\nu \right) $-Stirling's formula is computed as
the following form:

\begin{theorem}
($\left( \mu ,\nu \right) $-Stirling's formula) Let $n!_{\left( \mu ,\nu
\right) }$ be the $\left( \mu ,\nu \right) $-factorial defined by (\ref{def
of munu-kaijyo}). The $\left( \mu ,\nu \right) $-Stirling's formula $\ln
_{\mu }\left( n!_{\left( \mu ,\nu \right) }\right) $ is computed as follows:%
\begin{equation}
\ln _{\mu }\left( n!_{\left( \mu ,\nu \right) }\right) =\left\{ 
\begin{array}{ll}
\dfrac{n\ln _{\mu }n^{\nu }-\nu n}{\nu \left( 1-\mu \right) +1}+O\left( \ln
_{\mu }n\right) \quad  & \text{if}\quad \nu \left( 1-\mu \right) +1\neq 0,
\\ 
\nu \left( n-\ln n\right) +O\left( 1\right)  & \text{if}\quad \nu \left(
1-\mu \right) +1=0.%
\end{array}%
\right.   \label{munu_Stirling's formula}
\end{equation}
\end{theorem}

This formula is computed by the approximation:%
\begin{equation}
\ln _{q}\left( n!_{\left( \mu ,\nu \right) }\right) =\sum_{k=1}^{n}\ln _{\mu
}k^{\nu }\simeq \int_{1}^{n}\ln _{\mu }x^{\nu }dx.
\end{equation}

Based on these results, we obtain the one-to-one correspondence between the $%
\left( \mu ,\nu \right) $-multinomial coefficient and Tsallis entropy as
follows.

\begin{theorem}
\label{main result}When $n$ is sufficiently large, the $\mu $-logarithm of
the $\left( \mu ,\nu \right) $-multinomial coefficient coincides with
Tsallis entropy (\ref{Tsallis entropy}) as follows:%
\begin{equation}
\frac{1}{\nu }\ln _{\mu }\left[ 
\begin{array}{ccc}
& n &  \\ 
n_{1} & \cdots & n_{k}%
\end{array}%
\right] _{\left( \mu ,\nu \right) }\simeq \left\{ 
\begin{array}{ll}
\dfrac{n^{q}}{q}\cdot S_{q}\left( \dfrac{n_{1}}{n},\cdots ,\dfrac{n_{k}}{n}%
\right) & \text{if}\quad q\neq 0 \\ 
-S_{1}\left( n\right) +\sum\limits_{i=1}^{k}S_{1}\left( n_{i}\right) & \text{%
if}\quad q=0%
\end{array}%
\right.  \label{g-correspondence}
\end{equation}%
where $\nu \neq 0$,%
\begin{equation}
\nu \left( 1-\mu \right) +1=q,  \label{munuq-relation}
\end{equation}%
$S_{q}$ is Tsallis entropy (\ref{Tsallis entropy}) and $S_{1}\left( n\right)
:=\ln n.$
\end{theorem}

The proof is given in the appendix A.

The generalized correspondence (\ref{g-correspondence}) between the $\left(
\mu ,\nu \right) $-multinomial coefficient and Tsallis entropy includes some
typical mathematical structures such as the two kinds of dualities and the $%
q $-triplet as the special cases, shown in the next section.

\section{Multiplicative duality and other typical mathematical structures
derived from the combinatorial formalism}

We consider the case $q\neq 0$ only. Then, the generalized correspondence (%
\ref{g-correspondence}) is given by%
\begin{equation}
\frac{1}{\nu }\ln _{\mu }\left[ 
\begin{array}{ccc}
& n &  \\ 
n_{1} & \cdots  & n_{k}%
\end{array}%
\right] _{\left( \mu ,\nu \right) }\simeq \frac{n^{q}}{q}\cdot S_{q}\left( 
\dfrac{n_{1}}{n},\cdots ,\dfrac{n_{k}}{n}\right) .
\label{q-g-correspondence}
\end{equation}%
In this paper, we call the above relation (\ref{munuq-relation})
\textquotedblleft $\left( \mu ,\nu ,q\right) $\textit{-relation}%
\textquotedblright\ which provides us interesting features in Tsallis
statistics. In particular, we consider the following four cases.

\subsection{$\protect\nu =1$}

In this case, from the $\left( \mu ,\nu ,q\right) $-relation (\ref%
{munuq-relation}) $\mu $ is given by 
\begin{equation}
\mu =2-q.
\end{equation}%
Then the generalized correspondence (\ref{q-g-correspondence}) becomes%
\begin{equation}
\ln _{2-q}\left[ 
\begin{array}{ccc}
& n &  \\ 
n_{1} & \cdots  & n_{k}%
\end{array}%
\right] _{2-q}\simeq \frac{n^{q}}{q}\cdot S_{q}\left( \dfrac{n_{1}}{n}%
,\cdots ,\dfrac{n_{k}}{n}\right) 
\end{equation}%
which is equivalent to (\ref{rev_q-correspondence}) or (\ref%
{2-q-correspondence}) revealing the \textit{additive duality}
\textquotedblleft $q\leftrightarrow 2-q$\textquotedblright .

\subsection{$\protect\nu =q$}

In this case, from the $\left( \mu ,\nu ,q\right) $-relation (\ref%
{munuq-relation}) $\mu $ is determined as 
\begin{equation}
\mu =\frac{1}{q}.
\end{equation}%
Then the generalized correspondence (\ref{q-g-correspondence}) becomes%
\begin{equation}
\ln _{\frac{1}{q}}\left[ 
\begin{array}{ccc}
& n &  \\ 
n_{1} & \cdots & n_{k}%
\end{array}%
\right] _{\left( \frac{1}{q},q\right) }\simeq n^{q}\cdot S_{q}\left( \dfrac{%
n_{1}}{n},\cdots ,\dfrac{n_{k}}{n}\right)  \label{multiplicative duality0}
\end{equation}%
which reveals the \textit{multiplicative duality} \textquotedblleft $%
q\leftrightarrow \frac{1}{q}$\textquotedblright .

Aside from the above representation (\ref{multiplicative duality0}), the
multiplicative duality in Tsallis statistics is easily derived from the
definition of the escort distribution. See the appendix B for the detail.

\subsection{$\protect\nu =2-q$}

In this case, from the $\left( \mu ,\nu ,q\right) $-relation (\ref%
{munuq-relation}) $\mu $ is obtained as 
\begin{equation}
\mu =\frac{3-2q}{2-q}.  \label{mu_q}
\end{equation}%
Then the generalized correspondence (\ref{q-g-correspondence}) becomes%
\begin{equation}
\frac{1}{2-q}\ln _{\frac{3-2q}{2-q}}\left[ 
\begin{array}{ccc}
& n &  \\ 
n_{1} & \cdots & n_{k}%
\end{array}%
\right] _{\left( \frac{3-2q}{2-q},2-q\right) }\simeq \frac{n^{q}}{q}\cdot
S_{q}\left( \dfrac{n_{1}}{n},\cdots ,\dfrac{n_{k}}{n}\right)
\end{equation}%
where the $\left( \mu ,\nu ,q\right) $-relation for this case is equivalent
to the $q$\textit{-triplet} $\left( q_{\text{sen}},q_{\text{rel}},q_{\text{%
stat}}\right) $ recently conjectured by Tsallis \cite{Ts04PA}\cite{TGS05} in
the following sense. In \cite{Ts04PA}, Tsallis first conjectured the three
entropic $q$-indices $\left( q_{\text{sen}},q_{\text{rel}},q_{\text{stat}%
}\right) $, respectively for $q$-exponential \textit{sensitivity} to the
initial conditions, $q$-exponential \textit{relaxation} of macroscopic
quantities to thermal equilibrium and $q$-exponential distribution
describing a \textit{stationary} state. More concretely, based on his recent
results in \cite{MTG05}\ he conjectured the concrete $q$\textit{-triplet} $%
\left( q_{\text{sen}},q_{\text{rel}},q_{\text{stat}}\right) $ satisfying the
following relation \cite{TGS05}:%
\begin{equation}
q_{\text{rel}}+\frac{1}{q_{\text{sen}}}=2,\quad q_{\text{stat}}+\frac{1}{q_{%
\text{rel}}}=2.  \label{conjectured relation}
\end{equation}%
From this relation, we immediately obtain%
\begin{equation}
\frac{1}{q_{\text{sen}}}=\frac{3-2q_{\text{stat}}}{2-q_{\text{stat}}}
\end{equation}%
which is the \textit{same form} as $\mu $ obtained in (\ref{mu_q}).
Therefore, when $\nu =2-q$, the present $\left( \mu ,\nu ,q\right) $%
-relation is identified with the $q$-triplet $\left( q_{\text{sen}},q_{\text{%
rel}},q_{\text{stat}}\right) $ in the following sense: 
\begin{equation}
\mu =\frac{1}{q_{\text{sen}}},\text{\quad }\nu =\frac{1}{q_{\text{rel}}},%
\text{\quad }q=q_{\text{stat}}.  \label{identication}
\end{equation}%
As shown in this paper, the above identification (\ref{identication}) is
derived from the mathematical discussion \textit{only}. Besides our
analytical derivation, the $q$-triplet $\left( q_{\text{sen}},q_{\text{rel}%
},q_{\text{stat}}\right) $ has been already confirmed in the experimental
observations in \cite{BV05}. Therefore, Tsallis' conjecture on the $q$%
-triplet $\left( q_{\text{sen}},q_{\text{rel}},q_{\text{stat}}\right) $ in 
\cite{TGS05} is \textit{correct} in both theoretical and experimental
aspects.

Note that in our theoretical derivation of (\ref{identication}) we \textit{%
never} use the definition of the three entropic $q$-indices $q_{\text{sen}%
},q_{\text{rel}},q_{\text{stat}}$, which may be remained as a future work
from the theoretical points of view in this case $\nu =2-q$. However, the
present identification (\ref{identication}) is just an interpretation of the 
$\left( \mu ,\nu ,q\right) $-relation. In fact, in our formulation the
additive duality, the multiplicative duality and\textit{\ }the $q$-triplet
are derived as \textit{special cases} of the $\left( \mu ,\nu ,q\right) $%
-relation\ (\ref{munuq-relation}). For example, we present other possible
interpretation of the present $\left( \mu ,\nu ,q\right) $-relation\ (\ref%
{munuq-relation}) for the case $\nu =\frac{1}{q}$, shown in the next
subsection.

\subsection{$\protect\nu =\dfrac{1}{q}$}

As an other special case of the $\left( \mu ,\nu ,q\right) $-relation, we
consider the case $\nu =\frac{1}{q}$. For this case, we obtain%
\begin{equation}
\frac{1}{1-\mu }=\frac{1}{q-1}-\frac{1}{q}.  \label{result_case_D}
\end{equation}%
This identity reminds us of the following relationship \cite{LT98}:%
\begin{equation}
\frac{1}{1-q_{\text{sen}}}=\frac{1}{\alpha _{\min }}-\frac{1}{\alpha _{\max }%
}  \label{98LT_result}
\end{equation}%
where $q_{\text{sen}}$ is the same entropic $q$-index as the case C for the $%
q$-exponential \textit{sensitivity} to the initial conditions, $\alpha
_{\min }$ and $\alpha _{\max }$ are the values of $\alpha $ at which the
multifractal singularity spectrum $f\left( \alpha \right) $ vanishes (with $%
\alpha _{\min }<\alpha _{\max }$). These $\alpha _{\min }$ and $\alpha
_{\max }$ are given by%
\begin{equation}
\alpha _{\min }=\frac{\ln b}{z\ln \alpha _{F}},\quad \alpha _{\max }=\frac{%
\ln b}{\ln \alpha _{F}}
\end{equation}%
where $b$ stands for a natural scale for the partitions, $\alpha _{F}$ is
the Feigenbaum universal scaling factor and $z$ represents the nonlinearity
of the map at the vicinity of its extremal point \cite{LT98}. A choice of
the nonlinearity $z$ to satisfy%
\begin{equation}
\left( \frac{b}{\alpha _{F}}\right) ^{z}=b  \label{z_requirement}
\end{equation}%
implies $\alpha _{\max }-\alpha _{\min }=1$. In other words, (\ref%
{z_requirement}) means a rescaling of $\alpha _{\max }-\alpha _{\min }$ to
be $1$. Therefore, if the nonlinearity $z$ is determined by the above
requirement (\ref{z_requirement}), we have the following identification:%
\begin{equation}
\mu =q_{\text{sen}},\text{\quad }\nu =\frac{1}{\alpha _{\max }},\text{\quad }%
q=\alpha _{\max }  \label{identification_D}
\end{equation}%
which is one of the interpretations of the $\left( \mu ,\nu ,q\right) $%
-relation. Note that the above identification (\ref{identification_D})
implies (\ref{z_requirement}).

All results in cases A-D mean that the $\left( \mu ,\nu ,q\right) $-relation
is found to be a more general nature in Tsallis statistics to recover these
specific mathematical structures.

\section{Conclusion}

We present the one-to-one correspondence between the $\left( \mu ,\nu
\right) $-multinomial coefficient and Tsallis entropy $S_{q}$ to represent
both the additive duality \textquotedblleft $q\leftrightarrow 2-q$%
\textquotedblright\ and the multiplicative duality \textquotedblleft $%
q\leftrightarrow 1/q$\textquotedblright\ in one unified formula (\ref%
{q-g-correspondence}). In this derivation, $\left( \mu ,\nu \right) $%
-factorial, $\left( \mu ,\nu \right) $-multinomial coefficient and $\left(
\mu ,\nu \right) $-Stirling's formula are concretely formulated as a
generalization of $q$-factorial, $q$-multinomial coefficient and $q$%
-Stirling's formula, respectively. In the present one-to-one correspondence (%
\ref{q-g-correspondence}), when $\nu =2-q$, the $\left( \mu ,\nu ,q\right) $%
-relation among three parameters $\mu ,\nu ,q$ is shown to be identified
with the $q$-triplet $\left( q_{\text{sen}},q_{\text{rel}},q_{\text{stat}%
}\right) $ in the sense of (\ref{identication}). In addition, as other
interpretation of the $\left( \mu ,\nu ,q\right) $-relation, the
multifractal structure $1/\left( 1-q_{\text{sen}}\right) =1/\alpha _{\min
}-1/\alpha _{\max }$ is recoverd.

\bigskip 

\bigskip 

\textbf{Acknowledgement} \textit{The first author is grateful to Jan Naudts
for a short discussion in Trieste conference 2006, which inspires the author
to find the ideas in this paper. The authors acknowledge the partial support
given by the Ministry of Education, Science, Sports and Culture,
Grant-in-Aid for Scientific Research (B), 18300003, 2006.}

\appendix

\section{\textbf{Proof of Theorem 11}}

When $\mu =1$, $\mu $-product recovers the usual product regardless of $\nu $%
. Thus, we consider the case $\mu \neq 1$ only. 
\begin{align}
& \text{if}\quad \nu \left( 1-\mu \right) +1\neq 0,  \notag \\
& \ln _{\mu }\left[ 
\begin{array}{ccc}
& n &  \\ 
n_{1} & \cdots  & n_{k}%
\end{array}%
\right] _{\left( \mu ,\nu \right) }=\ln _{\mu }n!_{\left( \mu ,\nu \right)
}-\ln _{\mu }n_{1}!_{\left( \mu ,\nu \right) }-\cdots -\ln _{\mu
}n_{k}!_{\left( \mu ,\nu \right) } \\
& \simeq \frac{n\ln _{\mu }n^{\nu }-\nu n}{\nu \left( 1-\mu \right) +1}-%
\frac{n_{1}\ln _{\mu }n_{1}^{\nu }-\nu n_{1}}{\nu \left( 1-\mu \right) +1}%
-\cdots -\frac{n_{k}\ln _{\mu }n_{k}^{\nu }-\nu n_{k}}{\nu \left( 1-\mu
\right) +1}\quad \quad \left( \because \text{(\ref{munu_Stirling's formula})}%
\right)  \\
& =\frac{n\ln _{\mu }n^{\nu }}{\nu \left( 1-\mu \right) +1}-\frac{n_{1}\ln
_{\mu }n_{1}^{\nu }}{\nu \left( 1-\mu \right) +1}-\cdots -\frac{n_{k}\ln
_{\mu }n_{k}^{\nu }}{\nu \left( 1-\mu \right) +1}\quad \quad \left( \because
n=\sum_{i=1}^{k}n_{i}\right)  \\
& =\frac{n\left( n^{\nu \left( 1-\mu \right) }-1\right) }{\left( 1-\mu
\right) \left( \nu \left( 1-\mu \right) +1\right) }-\frac{n_{1}\left(
n_{1}^{\nu \left( 1-\mu \right) }-1\right) }{\left( 1-\mu \right) \left( \nu
\left( 1-\mu \right) +1\right) }-\cdots -\frac{n_{k}\left( n_{k}^{\nu \left(
1-\mu \right) }-1\right) }{\left( 1-\mu \right) \left( \nu \left( 1-\mu
\right) +1\right) } \\
& =\frac{n^{\nu \left( 1-\mu \right) +1}}{\left( 1-\mu \right) \left( \nu
\left( 1-\mu \right) +1\right) }-\frac{n_{1}^{\nu \left( 1-\mu \right) +1}}{%
\left( 1-\mu \right) \left( \nu \left( 1-\mu \right) +1\right) }-\cdots -%
\frac{n_{k}^{\nu \left( 1-\mu \right) +1}}{\left( 1-\mu \right) \left( \nu
\left( 1-\mu \right) +1\right) } \\
& =\frac{n^{\nu \left( 1-\mu \right) +1}}{\left( 1-\mu \right) \left( \nu
\left( 1-\mu \right) +1\right) }\left( 1-\left( \frac{n_{1}}{n}\right) ^{\nu
\left( 1-\mu \right) +1}-\cdots -\left( \frac{n_{k}}{n}\right) ^{\nu \left(
1-\mu \right) +1}\right)  \\
& =\frac{n^{\nu \left( 1-\mu \right) +1}}{\left( 1-\mu \right) \left( \nu
\left( 1-\mu \right) +1\right) }\left( 1-\sum\limits_{i=1}^{k}\left( \frac{%
n_{i}}{n}\right) ^{\nu \left( 1-\mu \right) +1}\right) 
\end{align}%
\begin{align}
& =\frac{\nu n^{\nu \left( 1-\mu \right) +1}}{\nu \left( 1-\mu \right) +1}%
\left( \frac{1-\sum\limits_{i=1}^{k}\left( \frac{n_{i}}{n}\right) ^{\nu
\left( 1-\mu \right) +1}}{\nu \left( 1-\mu \right) }\right) \qquad \qquad
\qquad \qquad \qquad \qquad \qquad \qquad \qquad  \\
& =\nu \frac{n^{\nu \left( 1-\mu \right) +1}}{\nu \left( 1-\mu \right) +1}%
S_{\nu \left( 1-\mu \right) +1}\left( \dfrac{n_{1}}{n},\cdots ,\dfrac{n_{k}}{%
n}\right) 
\end{align}%
\begin{align}
& \text{if}\quad \nu \left( 1-\mu \right) +1=0,  \notag \\
& \ln _{\mu }\left[ 
\begin{array}{ccc}
& n &  \\ 
n_{1} & \cdots  & n_{k}%
\end{array}%
\right] _{\left( \mu ,\nu \right) }=\ln _{\mu }n!_{\left( \mu ,\nu \right)
}-\ln _{\mu }n_{1}!_{\left( \mu ,\nu \right) }-\cdots -\ln _{\mu
}n_{k}!_{\left( \mu ,\nu \right) } \\
& \simeq \nu \left( n-\ln n\right) -\nu \left( n_{1}-\ln n_{1}\right)
-\cdots -\nu \left( n_{k}-\ln n_{k}\right) \quad \quad \left( \because \text{%
(\ref{munu_Stirling's formula})}\right)  \\
& =\nu \left( -\ln n+\ln n_{1}\cdots +\ln n_{k}\right) \quad \quad \left(
\because n=\sum_{i=1}^{k}n_{i}\right)  \\
& =\nu \left( -S_{1}\left( n\right) +\sum_{i=1}^{k}S_{1}\left( n_{i}\right)
\right) 
\end{align}

\section{\textbf{The} multiplicative duality derived from the definition of
the escort distribution}

The escort distribution was first introduced in \cite{BS93} to scan the
structure of a given distribution by using similarities as thermodynamic
equilibrium distribution.

\begin{definition}
(escort distribution) For any given probability distribution $\left\{
p_{i}\right\} $, the escort distribution $\left\{ P_{i}\right\} $ is defined
as 
\begin{equation}
P_{i}:=\frac{p_{i}^{q}}{\sum\limits_{j=1}^{n}p_{j}^{q}}\quad \left(
q>0\right) .  \label{def of escort}
\end{equation}
\end{definition}

Note that a given distribution $\left\{ p_{i}\right\} $ in the above
definition is \textit{not necessarily} normalized, but in our formulations
we require $\left\{ p_{i}\right\} $ to be a normalized distribution, that
is, a probability distribution.

Until now, the multiplicative duality \textquotedblleft $q\leftrightarrow
1/q $\textquotedblright\ in Tsallis statistics is based on the following
property derived from the above definition of the escort distribution:%
\begin{equation}
p_{i}=\frac{P_{i}^{\frac{1}{q}}}{\sum\limits_{j=1}^{n}P_{j}^{\frac{1}{q}}}.
\end{equation}%
However, the escort distribution $\left\{ P_{i}\right\} $ is originally
associated with Tsallis entropy in the following sense:

\begin{theorem}
For any given probability distribution $\left\{ p_{i}\right\} $ and its
associated escort distribution $\left\{ P_{i}\right\} $, the next identity
is satisfied. 
\begin{equation}
\exp _{q}\left( S_{q}\left( p_{i}\right) \right) =\exp _{\frac{1}{q}}\left(
S_{\frac{1}{q}}\left( P_{i}\right) \right)  \label{equivalence_0}
\end{equation}%
where $S_{q}\left( p_{i}\right) $ is Tsallis entropy (\ref{Tsallis entropy}).
\end{theorem}

\begin{proof}
Using the definition of the escort distribution (\ref{def of escort}), we
have%
\begin{equation}
\sum\limits_{i=1}^{n}P_{i}^{\frac{1}{q}}=\sum\limits_{i=1}^{n}\left( \frac{%
p_{i}^{q}}{\sum\limits_{j=1}^{n}p_{j}^{q}}\right) ^{\frac{1}{q}%
}=\sum\limits_{i=1}^{n}\frac{p_{i}}{\left(
\sum\limits_{j=1}^{n}p_{j}^{q}\right) ^{\frac{1}{q}}}=\left(
\sum\limits_{j=1}^{n}p_{j}^{q}\right) ^{-\frac{1}{q}}.
\end{equation}%
Both sides to the power $\frac{1}{1-\frac{1}{q}}=-\frac{q}{1-q}$ is 
\begin{equation}
\left( \sum\limits_{j=1}^{n}P_{j}^{\frac{1}{q}}\right) ^{\frac{1}{1-\frac{1}{%
q}}}=\left( \sum\limits_{i=1}^{n}p_{i}^{q}\right) ^{\frac{1}{1-q}}
\end{equation}%
which is equivalent to%
\begin{equation}
\left( 1+\left( 1-\frac{1}{q}\right) \frac{1-\sum\limits_{j=1}^{k}P_{j}^{%
\frac{1}{q}}}{\frac{1}{q}-1}\right) ^{\frac{1}{1-\frac{1}{q}}}=\left(
1+\left( 1-q\right) \frac{1-\sum\limits_{i=1}^{k}p_{i}^{q}}{q-1}\right) ^{%
\frac{1}{1-q}}.
\end{equation}%
Clearly, this is identical to the simple form%
\begin{equation}
\exp _{\frac{1}{q}}\left( S_{\frac{1}{q}}\left( P_{j}\right) \right) =\exp
_{q}\left( S_{q}\left( p_{i}\right) \right) .
\end{equation}
\end{proof}

Note that the above result (\ref{equivalence_0}) obviously reveals the
multiplicative duality \textquotedblleft $q\leftrightarrow \frac{1}{q}$%
\textquotedblright\ of Tsallis entropy, which is derived from the definition
of the escort distribution only.

The above relation (\ref{equivalence_0}) provides us a key to unify several
entropies such as Boltzmann-Gibbs-Shannon entropy, R\'{e}nyi entropy \cite%
{Re60}, Tsallis entropy \cite{Ts88}, Gaussian entropy \cite{FD00},
Sharma-Mittal entropy \cite{SM75} and Supra-extensive entropy \cite{Ma05}.
Among these entropies, Sharma-Mittal entropy and Supra-extensive entropy are
the two-parameterized entropies including the other entropies as special
cases. For these two entropies, the similar identities as above are
satisfied in the forms:%
\begin{eqnarray}
\exp _{r}\left( S_{q,r}^{\text{Sharma-Mittal}}\right) &=&\exp _{q}\left(
S_{q}^{\text{Tsallis}}\right) ,  \label{Sharma-Mittal} \\
\exp _{q}\left( S_{q,r}^{\text{Supra-extensive}}\right) &=&\exp _{r}\left(
S_{q}^{\text{R\'{e}nyi}}\right)  \label{Supra-extensive}
\end{eqnarray}%
where%
\begin{eqnarray}
S_{q,r}^{\text{Sharma-Mittal}}\left( p_{i}\right) &:&={\frac{\left( {{%
\sum\limits_{i=1}^{n}{p_{i}^{q}}}}\right) ^{\frac{1-r}{1-q}}-1}{{1-r}},} \\
S_{q,r}^{\text{Supra-extensive}}\left( p_{i}\right) &:&={\frac{\left( 1+%
\frac{1-r}{1-q}\log {{\sum\limits_{i=1}^{n}{p_{i}^{q}}}}\right) ^{\frac{1-q}{%
1-r}}-1}{{1-q}}.}
\end{eqnarray}%
The above identities (\ref{Sharma-Mittal}) and (\ref{Supra-extensive}) are
respectively simple mathematical modifications of (2.9) and (2.10) in \cite%
{Ma05}.

\end{document}